\documentclass{aa} 
 
\usepackage{graphicx}
\usepackage{multirow}
\usepackage{booktabs}
\usepackage{hyperref}
\usepackage[flushleft]{threeparttable}
\usepackage{txfonts}
\usepackage{natbib}

\usepackage[version=4]{mhchem}

\bibpunct{(}{)}{;}{a}{}{,}

\newcommand\reaction[1] {\begin{equation}\ce{#1}\end{equation}}

\begin{document}

   \title{Complex organic molecules in protoplanetary disks: X-ray photodesorption from methanol-containing ices \\\vspace{0.5cm} \Large Part II - Mixed methanol-CO and methanol-H$_2$O ices} 

   \author{     R. Basalg\`{e}te, \inst{1}    
            R. Dupuy, \inst{1}
            G. F\'{e}raud, \inst{1}
            C. Romanzin, \inst{2}
            L. Philippe, \inst{1}
            X. Michaut, \inst{1}
            J. Michoud, \inst{1}
            L. Amiaud, \inst{3}
            A. Lafosse, \inst{3}
            J.-H. Fillion,  \inst{1}
            M. Bertin \inst{1}                               
                  }

     \offprints{romain.basalgete@obspm.fr}

         \institute {$^1$ Sorbonne Universit\'{e}, Observatoire de Paris, PSL University, CNRS, LERMA, F-75014, Paris, France \\
         $^2$ Univ Paris Saclay, CNRS UMR 8000, ICP, F-91405, Orsay, France \\
         $^3$ Univ Paris Saclay, CNRS, ISMO, F-91405, Orsay, France}
         
     
     \date{Received 11 December 2020 ; Accepted 5 January 2021}

 \abstract{Astrophysical observations show complex organic molecules (COMs) in the gas phase of protoplanetary disks. X-rays emitted from the central young stellar object (YSO) that irradiate interstellar ices in the disk, followed by the ejection of molecules in the gas phase, are a possible route to explain the abundances observed in the cold regions. This process, known as X-ray photodesorption, needs to be quantified for methanol-containing ices.}{We aim at experimentally measuring X-ray photodesorption yields (in molecule desorbed per incident photon, displayed as molecule/photon for more simplicity) of methanol and its photo-products from binary mixed ices: $^{13}$CO:CH$_3$OH ice and H$_2$O:CH$_3$OH ice.}{We irradiated these ices at 15 K with X-rays in the 525 - 570 eV range from the SEXTANTS beam line of the SOLEIL synchrotron facility. The release of species in the gas phase was monitored by quadrupole mass spectrometry, and photodesorption yields were derived.}{For $^{13}$CO:CH$_3$OH ice,  CH$_3$OH X-ray photodesorption yield is estimated to be $\sim 10^{-2}$ molecule/photon at 564 eV. X-ray photodesorption of larger COMs, which can be attributed to either ethanol, dimethyl ether, and/or formic acid, is detected with a yield of $\sim 10^{-3}$ molecule/photon. When methanol is mixed with water, X-ray photodesorption of methanol and of the previous COMs is not detected. X-ray induced chemistry, dominated by low-energy secondary electrons, is found to be the main mechanism that explains these results. We also provide desorption yields that are applicable to protoplanetary disk environments for astrochemical models.}{The X-ray emission from YSOs should participate in the enrichment of the protoplanetary disk gas phase with COMs such as methanol in the cold and X-ray dominated regions because of X-ray photodesorption from methanol-containing ices.} 

   
             
   \titlerunning{X-ray photodesorption from methanol-containing ices}  
   \authorrunning{Basalg\`ete et al.}  
 
   \maketitle

\section{Introduction} 

Methanol (CH$_3$OH) is a species of prime importance in astrochemistry. It is an organic molecule  that is commonly referred to as a complex organic molecule (COM) and has been detected in many regions of star and planet formation in the interstellar medium (e.g., \cite{guzman2013, vastel2014}). It is thought to play a central role in the chemical evolution of these media, potentially leading to the production of more complex species \citep{garrod_complex_2008, elsila_mechanisms_2007}. In protoplanetary disks, its detection in the gas phase has been confirmed \citep{walsh_first_2016, carney_upper_2019}, together with other oxygen-bearing COMs such as formic acid (HCOOH) or formaldehyde (H$_2$CO) \citep{favre_first_2018, podio_organic_2019}.
\\\\
At the low temperature in the regions where it is detected, methanol is believed to be mainly accreted onto the surface of the dust grains. So far, methanol is the only COM that has been shown by mid-infrared detection to be a constituent of the icy mantles. Its presence in the ices has been highlighted at several stages of star formation  \citep{taban_stringent_2003, gibb_interstellar_2004, boogert_observations_2015}. Its concentration varies between 1 to 25 \% with respect to water. The lack of known efficient ways for methanol formation in the gas phase has led to the general assumption that it is solely synthesized in the icy mantles, although this is still debated \citep{dartois_non-thermal_2019}. Pioneer experimental studies have shown that its formation can proceed through successive H-addition in the CO-rich upper layers of the ices beyond the CO snow lines \citep{watanabe_efficient_2002}, resulting in methanol embedded in a CO-rich environment. Nevertheless, other chemical routes were also invoked in the condensed phase. \cite{qasim_formation_2018} showed that methanol can also be formed, although with a lower efficiency, from sequential reactions involving methane in CO-poor and H$_2$O-rich ices, demonstrating that methanol could be present in the condensed phase at an earlier stage of the ice evolution. 
\\\\
Because methanol most likely originates in the ice mantle, nonthermal desorption pathways have to be invoked to explain its observation in the gas phase in the cold regions where thermal desorption is not operative. In the past years, several experimental studies were conducted to highlight and quantify several nonthermal processes. The effect of cosmic rays on ices comprising methanol has been shown to be a possible route of bringing intact methanol molecules into the gas phase \citep{dartois_cosmic_2018}. The chemical desorption, that is, the desorption of methanol following its exothermic formation onto the grain surface, has also been studied, but no experimental evidence of this process has been gained when the reaction takes place on CO ice surfaces \citep{minissale_hydrogenation_2016}. Vacuum-ultraviolet (VUV) photodesorption, that is, the desorption triggered by the effect of VUV photons (5-13,6 eV) on the ices, has also been shown to be able to desorb methanol molecules from pure methanol ices, with yields of about $10^{-5}$ ejected molecules per incident photon \citep{bertin_uv_2016, cruz-diaz_negligible_2016}. However, this process has also been shown to depend on the ice composition. From model ices of more relevant composition, that is, methanol condensed into a CO-rich matrix, no methanol desorption could be measured, and only an upper value of $\sim 10^{-6}$ desorbed methanol molecules per incident photon could be established \citep{bertin_uv_2016}, which was explained by the methanol desorption being a consequence of the VUV photochemistry in the solid phase.
\\\\
Recently, another nonthermal desorption process has experimentally been shown to be an efficient way for maintaining molecules in the gas phase of the cold regions of the interstellar medium: the desorption induced by X-rays in the spectral range $0,5 - 10$ keV, which was first studied in the case of water ices, and has been shown to be at least competitive with the VUV photodesorption in protoplanetary disks \citep{dupuy_x-ray_2018}. Young stellar objects (YSOs) (Class I, Class II, and Class III) have been shown to be X-ray emitters in the range of $\sim$0.1-10 keV \citep{imanishi_2003, ozawa_x-ray_2005, giardino_onset_2007}, with a typical luminosity of $\sim10^{30}$ erg.s$^{-1}$. This X-ray radiation field can reach regions of protoplanetary disks that are shielded from VUV photons \citep{agundez_chemistry_2018, walsh_molecular_2015}, and can originate in the central star and also in other surrounding young stars inside an YSO cluster, thus irradiating the disk out of its plane \citep{adams_background_2012}.
\\\\
The role played by these X-rays in the nonthermal desorption of methanol in disks is so far an open question. In paper I, we have performed an experimental study of the methanol X-ray photodesorption from pure methanol ice. We showed that under the radiation conditions of protoplanetary disks, X-rays are at least as efficient as UV photons in desorbing methanol from pure methanol ice. Together with methanol desorption, we showed the desorption of other smaller species, and also the desorption of more complex molecules that could be associated with either formic acid, dimethyl ether, or ethanol. We proposed that the X-ray photodesorption is mainly carried out by the thermalization of Auger electrons produced by the X-ray photon absorption and ionization of the 1s electron of the oxygen atom, resulting in the production of a large number of secondary electrons in the ice. The methanol desorption much likely arises from a subsequent chemistry induced by these secondary electrons. When complex chemistry is involved, it is therefore expected that the desorption process strongly depends on the composition of the ice. This has previously been highlighted in the case of VUV photodesorption in ices in which CO and methanol are mixed. 
\\\\
As stated before, methanol is likely formed directly in the ice mantles, either by hydrogenation of the CO upper layers or in a more H$_2$O-rich phase. It appears to be unlikely that pure methanol phases exist in interstellar ices. The effect of the ice composition therefore needs to be evaluated, and more realistic yields need to be extracted from methanol-containing ices of more relevant compositions. This is the aim of this paper II, in which we describe the X-ray photodesorption from methanol-containing model ices that are frozen mixtures of methanol with CO and of methanol with H$_2$O. The experimental method is the same as we used in paper I, and the studies harness the high brilliance and accordability of the synchrotron radiation in the soft X-ray range that is provided by the SEXTANTS beam line
of the SOLEIL facility (St Aubin, France).
\\\\
Paper II is organized as follows. Section 2 summarizes the experimental method that is described in more detail in paper I. In section 3 we report the X-ray absorption profile of the ices, and we summarize the photodesorption yields derived from our experiments. In section 4 we discuss the identified relevant mechanisms involved in the photodesorption process regarding our data, and we extrapolate these results to study their astrophysical implications. This study is intimately linked to the discussions conducted in paper I, where we focused on the specific case of X-ray photodesorption from pure methanol ice. Some results are therefore directly taken from paper I in order to discuss the results obtained in the present study.

\section{Experimental procedures} 
Experiments were conducted using the SPICES 2 set-up (Surface Processes and ICES 2). It consists of an ultra-high vacuum (UHV) chamber (base pressure $\sim 10^{-10}$ mbar) within which a rotatable copper substrate (polycrystalline oxygen-free high-conductivity copper) can be cooled down to T$\sim$15 K by a closed-cycle helium cryostat. It is electrically insulated from its sample holder by a Kapton foil, allowing the measurement of the drain current that is generated by the electrons escaping from its surface after X-ray absorption. This is referred to as the total electron yield (TEY) in the following. On this substrate the ices  are formed using a dosing system that allows injecting a gaseous mixture with different stoichiometry factors of methanol (99.9\% purity, Sigma-Aldrich) and $^{13}$CO (99\% $^{13}$C purity, eurisotop) or methanol and H$_2$O (liquid chromatography standard, Fluka) directly on its surface. Gaseous mixtures of several methanol concentration are prepared in a gas-introduction system equipped with a capacitive pressure gauge before introduction into the experiment. The exact stoichiometries of the resulting ices are controlled using the temperature- programmed desorption (TPD) technique, with which ice thicknesses expressed in monolayer (ML), which is equivalent to a surface density of $\sim 10^{15}$ molecules/cm$^2$ , and composition are calibrated \citep{doronin_adsorption_2015}. The relative precision of this technique is about 10 \%. In our experiments, we studied ices of $\sim$100 ML and with $^{13}$CO:CH$_3$OH ratios of $\sim$1:1 and $\sim$6:1 and H$_2$O:CH$_3$OH ratios of $\sim$1.5:1 and $\sim$3:1. $^{13}$CO:CH$_3$OH ices were formed at 15 K. H$_2$O:CH$_3$OH ices were formed at 90 K and cooled down to 15 K before irradiation to ensure that the resulting water ice was in its compact amorphous phase. This phase is commonly referred to as compact amorphous solid water (c-ASW).
\\\\
The X-ray photon source of the SEXTANTS beam line of the SOLEIL synchrotron facility was connected to the SPICES 2 setup to run our experiments. We used photons in the range of 525-570 eV, corresponding to the ionization edge of the O(1s) electron of water, CO, and methanol, with a resolution of $\Delta E = 150$ meV, where $E$ is the photon energy. The flux, measured with a calibrated silicon photodiode, was approximately $1.5 \times 10^{13}$ photon.s$^{-1}$, with little variation except for a significant dip around 534 eV that is due to the O 1s absorption of oxygen pollution on the optics of the beam line. The beam was set at a $47^{\circ}$ incidence on the ice surface in a spot of $\sim$0.1 cm$^2$.
\\\\
While the ices were irradiated, the photodesorption of neutral species was monitored by recording the desorbed molecules in the gas phase using a quadripolar mass spectrometer (QMS). Different irradiation procedures were made : 
\\\\
- short irradiations at 534, 541 and 564 eV were conducted to measure the photodesorption yields at these fixed energies. The irradiation time per fixed energy is approximately 10 seconds (higher than the acquisition time of the QMS, which is 100 ms). This mode allowed us to probe the photodesorption from the ices with a relatively low irradiation fluence, mainly to prevent the photoaging of the condensed systems from significantly affecting the detected signals. The fluence received by the ice during these short irradiations was $\sim 5 \times 10^{15}$ photon/cm$^{2}$.
\\\\
- continuous irradiations from 525 to 570 eV by 0.5 eV steps (scans) allowed us to record the photodesorption spectra as a function of the photon energy. The irradiation time per scan was approximately 10 minutes, and the fluence received by the ice was $\sim 1 \times 10^{17}$ photon/cm$^{2}$. During these irradiations, the TEYs were also measured as a function of the photon energy with a scan step of 0.5 eV.
\\\\
Finally, at the end of these irradiation experiments, TPD experiments (from 15 K to 200 K) were used to evaporate all the molecules from the substrate surface before a new ice was formed. These TPD experiments allowed us to compute the stoichiometry factors for mixture ices reported before more precisely, assuming that the composition of the ice before TPD is not globally modified by X-ray absorption. This assumption is reasonable considering the irradiated volume ($\sim$100 ML $\times$ 0.1 cm$^2$) compared to the total volume of the ice ($\sim$100 ML $\times$ 2.25 cm$^2$).
\\\\
The photodesorption yields were derived (in molecule desorbed per incident photon, displayed as molecule/photon for more simplicity in the following) using the signal given by the QMS, corrected for the photon flux and the apparatus function of the QMS and by applying a proportionality factor between the molecular desorption flux and the raw signal. A more detailed explanation of the calibration procedure is given in paper I. For the binary mixed ices we studied, we corrected the photodesorption yields for the dilution factor of one of the two initial molecules of the binary mixed ice. This method allows us to adequately compare the molecular desorption flux between pure ice (from paper I) and binary mixed ices by normalizing to the average surface density of the molecules, assuming a homogeneous mixing in the solid phase. A more detailed explanation is given in section 3.2.

\section{Results} 
\subsection{X-ray absorption of the ices and TEY} 
\begin{figure} [h!]
\resizebox{\hsize}{!}{\includegraphics{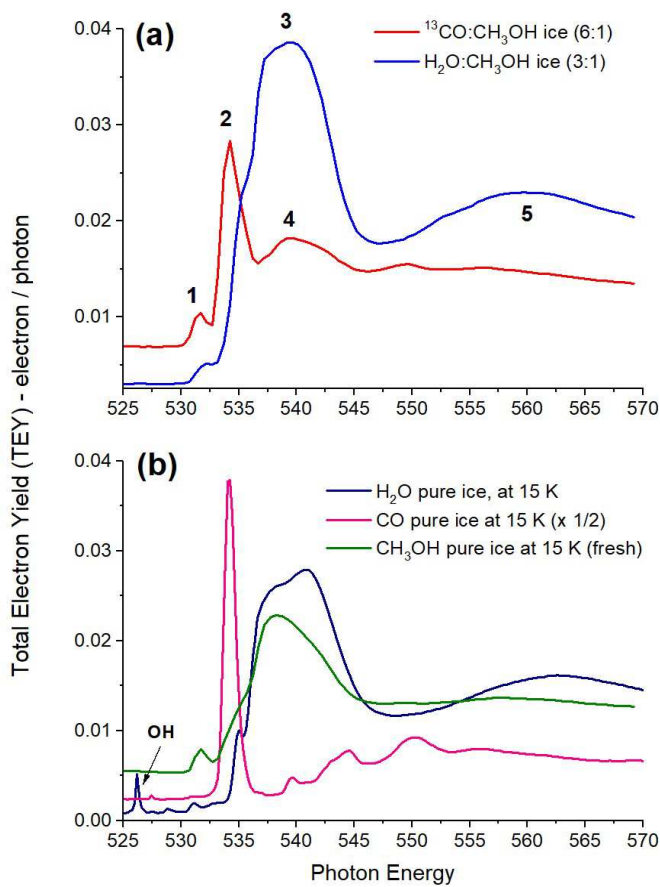}}
\caption{(a) TEY as a function of photon energy for $^{13}$CO:CH$_3$OH (6:1) and H$_2$O:CH$_3$OH (3:1) ices at 15 K ($\sim$100 ML). (b) TEY of pure CH$_3$OH (from paper I), pure CO and c-ASW H$_2$O ices measured at 15 K for $\sim$100 ML \citep{dupuy_x-ray_2018, dupuy:tel-02354689}}
\label{TEY}
\end{figure}

The TEYs measured from our experiments are shown in Figure \ref{TEY}.(a). The different features we observed are labeled and can be attributed to the X-ray absorption of the ice: after a photon is absorbed by a core O 1s electron, the decay of the resulting molecular excited state leads to the release of an Auger electron of $\sim$500 eV. The thermalization of this Auger electron by inelastic scattering within the ice creates secondary valence excitations and ionizations of neighboring molecules, leading to a cascade of secondary electrons. The current generated by the escape of these electrons from the ice surface can be quantified per incident photon and provides information about the core electronic structure of O-bearing molecules in condensed phase near the O K edge.\\

\begin{figure*} 
\centering
\includegraphics[width=17cm]{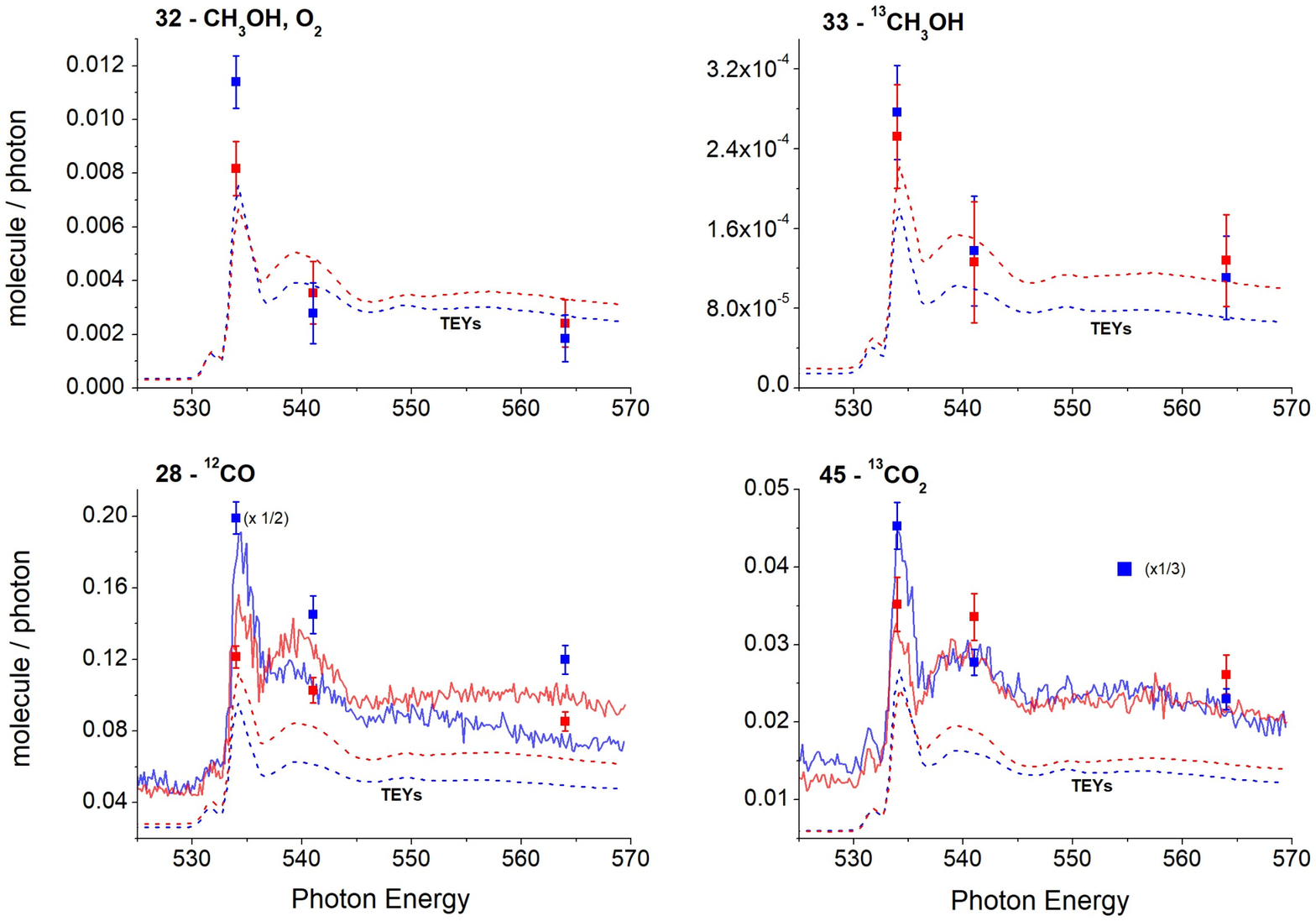}
\caption{Photodesorption spectra for masses 28, 32, 33, and 45 in molecule/photon from $^{13}$CO:CH$_3$OH ice at 15 K, with the associated molecules. Red and blue are associated to a mixing ratio of 1:1 and 6:1, respectively. The measurements at fixed energy are associated with a fluence before irradiation of $1 \times 10^{16}$ ph/cm$^2$ and are represented by the squares with error bars. The scan experiments are associated with a fluence before irradiation of $3 \times 10^{17}$ ph/cm$^2$ and are represented by solid lines. The TEYs measured during the scan experiments are also shown in arbitrary units by the dashed lines. The displayed photodesorption yields are not corrected for any dilution factor.}
\label{Graph_13CO}
\end{figure*}

\begin{figure*}
\centering
\includegraphics[width=17cm]{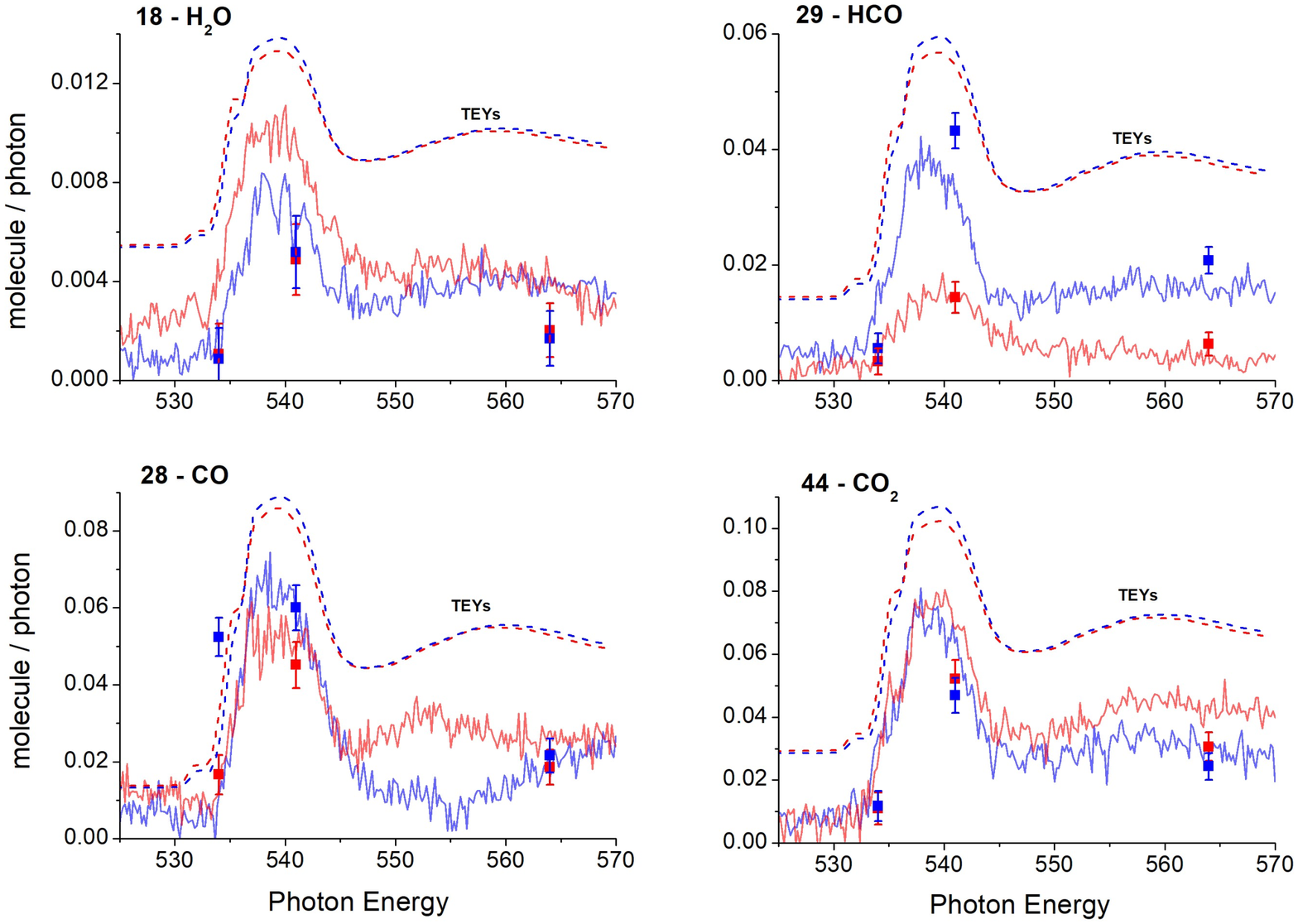}
\caption{Photodesorption spectra for masses 18, 28, 29, and 44 in molecule/photon from H$_2$O:CH$_3$OH ice at 15 K, with the associated molecules. Red and blue are associated with a mixing ratio of 1.5:1 and 3:1, respectively. The measurements at fixed energy are associated with a fluence before irradiation of $1 \times 10^{16}$ ph/cm$^2$ and are represented by the squares with error bars. The scan experiments are associated with a fluence before irradiation of $3 \times 10^{17}$ ph/cm$^2$ and are represented by solid lines. The TEYs measured during the scan experiments are also shown in arbitrary units by the dashed lines. The photodesorption yields displayed are not corrected for any dilution factor.}
\label{Graph_H2O}
\end{figure*}

In Figure \ref{TEY}.(b) we present the TEYs measured for pure CH$_3$OH (from paper I), pure CO, and pure H$_2$O ices at 15 K and for $\sim$100 ML \citep{dupuy_x-ray_2018, dupuy:tel-02354689}. When we compare Figure \ref{TEY}.(a) and Figure \ref{TEY}.(b), we clearly see that the features observed in the TEYs of our binary mixed ices (Figure \ref{TEY}.(a)) can be attributed to individual contributions of CH$_3$OH, CO, or H$_2$O as follows:
\\\\
- peak 1 was discussed in paper I, and we were unable to attribute it because we lack published data in the considered energy range.
\\
- the sharp peak 2 observed at 534.4 eV for the $^{13}$CO:CH$_3$OH ice can be attributed to the 1s$^{-1} \pi^*$ resonance (1$\sigma$-2$\pi$ transition) of CO molecules in condensed phase. This is also observed for CO in gas phase \citep{puttner_vibrationally_1999} and for pure CO ice (Figure \ref{TEY}.(b)).
\\
- the broadened peak 3 observed in H$_2$O:CH$_3$OH ice results from the overlap of CH$_3$OH and H$_2$O molecular absorption. For H$_2$O molecules, experiments on water core excitation attributed this feature to Rydberg orbitals \citep{tronc_role_2001, Parent_2002}.
\\
- the broadened peak 4 corresponds to CH$_3$OH 3p Rydberg orbitals, with some $\sigma^*$ character of the C-O bond (\citep{wilson_x-ray_2005}).
\\
- the broadened peak 5 corresponds to the first EXAFS oscillation. This is also observed in pure water ice \citep{dupuy_x-ray_2018, dupuy_desorption_2020}.
\\
- the structured features observed between 525 eV and 534 eV in pure water ice (Figure \ref{TEY}.(b)), resulting from X-ray induced chemistry are not seen in our H$_2$O:CH$_3$OH ice (Figure \ref{TEY}.(a)), especially for radical OH formation at 526 eV (identified by \citealt{Laffon_2006}). This point is discussed in section 4.

\subsection{Photodesorption yields}
In Figures \ref{Graph_13CO} and \ref{Graph_H2O} we report the photodesorption yields from $^{13}$CO:CH$_3$OH ice and H$_2$O:CH$_3$OH ice, respectively, in molecule desorbed per incident photon (displayed molecule/photon for more simplicity) as derived from our measurements. We do not display all the data available for more clarity because we did not observe differing behaviors from the presented data. The TEY measurements are also shown in arbitrary units (only the energy dependence is of interest when comparing with the photodesorption). The photodesorption yields derived from the irradiations at fixed energy are consistent with those measured during the scan experiments. The remaining relevant data we obtained are summarized in Tables \ref{Yields_Simple_Mol} and \ref{Yield_COMS}, where the yields are derived from our fixed energy experiments. The yields for pure CH$_3$OH ice (from paper I) are also displayed. As a lower fluence is used than in the scan experiments, the aging effect is limited for these yields: the fluence received by the ice before measurement ranges from $5 \times 10^{15}$ to $2 \times 10^{16}$ photon/cm$^{2}$. Moreover, these yields are derived at the fixed energy of 564 eV in order to avoid any resonance effect that might occur in the 535-545 eV region as a result of CH$_3$OH, CO and/or H$_2$O X-ray absorption when comparing the yields between the ices.
\\\\
In these tables and for $^{13}$CO:CH$_3$OH and H$_2$O:CH$_3$OH mixtures alone, we corrected the photodesorption yields for dilution factors depending on the molecule considered. This correction is necessary to adequately compare the photodesorption yields between pure and binary mixed ices. When we consider that the desorption of a given molecule from a binary mixed ice is due to the presence of one of the two molecules, which we call the parent molecule in the following discussion, we have to take into account that there are fewer parent molecules on the surface of the binary mixed ice than on the surface of the parent pure ice. The considered molecular desorption flux is therefore necessarily lower for the binary mixed ice than for the pure ice, which will make the comparison of the raw photodesorption yields irrelevant. The correction factor that should be considered corresponds to the fractional abundance of the parent molecule in the mixed ice. For example, $^{12}$CO$_2$ or $^{12}$CO photodesorption from $^{13}$CO:CH$_3$OH ice with a mixing ratio of 6:1 obviously originates from the presence of methanol in the ice. Thus, to be able to adequately compare these photodesorption yields with the ones from pure methanol ice, we should consider the yields per available methanol molecules at the surface, that is, multiplying the photodesorption yield for the binary mixed ice by 7.
\\\\
Accordingly, the photodesorption yields from the binary mixed $^{13}$CO:CH$_3$OH and H$_2$O:CH$_3$OH ices displayed in Tables \ref{Yields_Simple_Mol} and \ref{Yield_COMS} were corrected as follows: 
\\\\
- for $^{13}$CO:CH$_3$OH ices, we corrected $^{13}$CO and $^{13}$CO$_2$ yields for the dilution factor of $^{13}$CO (by multiplying by 2 and 7/6 for 1:1 and 6:1 mixing ratio, respectively), and we corrected the yields of all other molecules presented in Tables \ref{Yields_Simple_Mol} and \ref{Yield_COMS} for the dilution factor of CH$_3$OH (by multiplying by 2 and 7 for 1:1 and 6:1 mixing ratio, respectively).
\\
- for H$_2$O:CH$_3$OH ice, we corrected H$_2$O yields for the dilution factor of H$_2$O (by multiplying by 2.5/1.5 and 4/3 for 1.5:1 and 3:1 mixing ratio, respectively), and we corrected the yields of all other molecules presented in Tables \ref{Yields_Simple_Mol} and \ref{Yield_COMS} for the dilution factor of CH$_3$OH (by multiplying by 2.5 and 4 for 1.5:1 and 3:1 mixing ratio, respectively).

\begin{table*}[h!]
\caption{X-ray photodesorption yields at 564 eV of simple molecules from pure methanol ice (from paper I) and $^{13}$CO:CH$_3$OH and H$_2$O:CH$_3$OH mixtures, at 15 K and for a dose between $5 \times 10^{15}$ and $2 \times 10^{16}$ photon/cm$^{2}$. As explained in section 3.2, these yields were corrected for the dilution factor of the corresponding parent molecule for $^{13}$CO:CH$_3$OH and H$_2$O:CH$_3$OH ices.}
\begin{center}
\begin{threeparttable}
\centering
\begin{tabular}{p{5cm}p{2,1cm}p{2,1cm}p{2,1cm}p{2,1cm}p{2,1cm}}
\hline \hline \\
Photodesorption yields & &   
\multicolumn{2}{c}{\multirow{2}{*}{$^{13}$CO:CH$_3$OH ice}} &
\multicolumn{2}{c}{\multirow{2}{*}{H$_2$O:CH$_3$OH ice}}  \\
in $10^{-3}$ molecule / photon  \\\cmidrule{3-4}\cmidrule{5-6}\\
Mass - photodesorbed species & \centering Pure CH$_3$OH & \centering 1:1 & \centering 6:1 & \centering 1.5:1 & \centering 3:1 
\tabularnewline \hline\\

15 - CH$_3$\tnote{(3)} & \centering $1.3^{\pm0.2}$  & \centering $3.0^{\pm0.3}$ & \centering $7.1^{\pm0.7}$ & \centering $1.8^{\pm0.2}$ & \multicolumn{1}{c}{ $3.0^{\pm0.3}$ } \\ [0.2cm] 

16 - CH$_4$, O,$^{13}$CH$_3$ &  \centering $1.1^{\pm0.2}$ & \centering $3.8^{\pm0.5}$ & \centering $15^{\pm2}$  & \centering $1.4^{\pm0.2}$  &  \multicolumn{1}{c}{$2.0^{\pm0.3}$}  \\ [0.2cm]

17 - OH, $^{13}$CH$_4$\tnote{(1)} & \centering ND\tnote{**} & \centering $1.7^{\pm0.7}$ & \centering $7.9^{\pm3.2}$ & \centering ND\tnote{**} & \multicolumn{1}{c}{ND\tnote{**} } \\ [0.2cm]

18 - H$_2$O & \centering $7.5^{\pm1.1}$ & \centering NM\tnote{*} & \centering NM\tnote{*} & \centering $3.4^{\pm0.4}$ & \multicolumn{1}{c}{$2.3^{\pm0.4}$}  \\ [0.2cm] 

28 - CO &  \centering $(1.2)^{\pm0.1} \times 10^{2}$ & \centering $(1.7)^{\pm0.1} \times 10^{2}$ & \centering $(8.4)^{\pm0.2} \times 10^{2}$ & \centering $46^{\pm2}$ & \multicolumn{1}{c}{$86^{\pm4}$}  \\ [0.2cm]  

29 - HCO, $^{13}$CO & \centering NM\tnote{*} & \centering $(2.5)^{\pm0.1} \times 10^{2}$ & \centering $(4.4)^{\pm0.2} \times 10^{2}$ & \centering $16^{\pm1}$ & \multicolumn{1}{c}{$83^{\pm4}$}  \\ [0.2cm]

31 - CH$_2$OH, CH$_3$O &  \centering ND\tnote{(2)} & \centering ND\tnote{(2)} & \centering ND\tnote{(2)} & \centering $1.3^{\pm0.3}$ & \multicolumn{1}{c}{$5.2^{\pm1.0}$} \\ [0.2cm]

44 - CO$_2$ & \centering $24^{\pm1}$ &  \centering $85^{\pm4}$ & \centering $(2.6)^{\pm0.2} \times 10^{2}$ & \centering $77^{\pm4}$ & \multicolumn{1}{c}{$97^{\pm5}$} \\ [0.2cm] 

45 - $^{13}$CO$_2$ & & \centering $52^{\pm3}$ & \multicolumn{1}{c}{$80^{\pm4}$}
\\ [0.2cm]
\hline
\end{tabular}

\begin{tablenotes}
            \item[*] NM = Not measured: we did not measure the corresponding mass in the QMS for the ice considered 
            \item[**] ND = Not detected: the desorption signal measured for the considered species is below our signal-to-noise ratio, meaning that we did not detect its desorption from the ice. Considering the noise profile on the mass channel 17 of our QMS, if OH photodesorption occurs, the photodesorption yield is $< 5 \times 10^{-4}$ molecule/photon. 
              \item[(1)] For $^{13}$CO:CH$_3$OH mixtures only. The values given for the mass 17 are given without being able to correct from the cracking of possible photodesorbed H$_2$O into OH in the QMS. Thus these values may be overestimated.
               \item[(2)] We assumed that the desorption of CH$_2$OH and/or CH$_3$O from pure methanol ice and $^{13}$CO:CH$_3$OH mixtures is negligible.
                \item[(3)] These yields are given without being able to correct from the cracking of CH$_4$ into CH$_3$ in the QMS. Thus these values may be overestimated. 
\end{tablenotes}

\end{threeparttable}
\end{center}
\label{Yields_Simple_Mol}

\end{table*}

\begin{table*}
\caption{X-ray photodesorption yields at 564 eV of methanol and photoproducts from pure methanol ice (from paper I) and $^{13}$CO:CH$_3$OH and H$_2$O:CH$_3$OH mixture ices at 15 K and for a dose between $5 \times 10^{15}$ and $2 \times 10^{16}$ photon/cm$^{2}$. As explained in section 3.2, these yields are corrected for the dilution factor of the corresponding parent molecule for $^{13}$CO:CH$_3$OH and H$_2$O:CH$_3$OH ices.}
\begin{threeparttable}
\centering
\begin{tabular}{p{5,3cm}p{2,1cm}p{2,1cm}p{2,1cm}p{2,1cm}p{2,1cm}}
\hline \hline \\
Photodesorption yields & &   
\multicolumn{2}{c}{\multirow{2}{*}{\centering $^{13}$CO:CH$_3$OH ice}} &
\multicolumn{2}{c}{\multirow{2}{*}{\centering H$_2$O:CH$_3$OH ice}} \\
in $10^{-3}$ molecule / photon  \\\cmidrule{3-4}\cmidrule{5-6}\\
Mass - photodesorbed species & \centering Pure CH$_3$OH & \centering 1:1 & \centering 6:1 & \centering 1.5:1 & \centering 3:1 \tabularnewline \hline\\

30 - H$_2$CO & \centering $1.2^{\pm0.5}$  & \centering NM\tnote{*} & \centering NM\tnote{*} & \centering $1.1^{\pm0.5}$ & \multicolumn{1}{c}{$2.3^{\pm0.9}$} \\ [0,2cm]

31 - H$_2^{13}$CO\tnote{(1)} & & \centering $< 0.5$ & \multicolumn{1}{c}{\centering $8.8^{\pm1.8}$} \\ [0,2cm]

32 - CH$_3$OH\tnote{(1)} & \centering $7.6^{\pm0.9}$ & \centering $4.8^{\pm0.5}$ & \centering $13^{\pm2}$ & \centering ND\tnote{**} & \multicolumn{1}{c}{ND\tnote{**}} \\ [0,2cm]

33 - $^{13}$CH$_3$OH & & \centering $0.3^{\pm0.1}$ &\multicolumn{1}{c}{ $0.8^{\pm0.3}$} \\ [0,2cm]

46 - HCOOH, C$_2$H$_6$O isomers\tnote{(2)} & \centering $0.8^{\pm0.3}$ & \centering $1.5^{\pm0.8}$ & \centering $3.5^{\pm1.8}$ & \centering ND\tnote{**} & \multicolumn{1}{c}{\centering ND\tnote{**}} \\ [0,2cm] 

47 - H$^{13}$COOH, $^{13}$C$^{12}$CH$_6$O isomers & & \centering  $0.5^{\pm0.1}$ & \multicolumn{1}{c}{$4.2^{\pm0.8}$} \\ [0,2cm]
\hline
\end{tabular}

\begin{tablenotes}
            \item[*] NM = Not measured: we did not measure the corresponding mass in the QMS for the ice considered. 
            \item[**] ND = Not detected: the desorption signal measured for the considered species is below our signal-to-noise ratio, meaning that we did not detect its desorption from the ice. Considering the noise profile on the mass channel 32 and 46 of our QMS, if photodesorption from H$_2$O:CH$_3$OH ice occurs in these channels, the photodesorption yield is $<5 \times 10^{-4}$ molecule/photon.
            \item[(1)] These yields are estimated using the signal on the mass channel 31 (see section 3.2 for more details)
                \item[(2)] The photodesorption yields derived for the mass 46 on the $^{13}$CO:CH$_3$OH mixtures are not corrected for any possible cracking of the mass 47 in the QMS as no identification is possible for the molecules corresponding to the mass 47. 
\end{tablenotes}

\end{threeparttable}

\label{Yield_COMS}

\end{table*}

\subsubsection*{Simple molecules}
In Table \ref{Yields_Simple_Mol} we present the photodesorption yields of simple molecules at 564 eV and for a fluence between $5 \times 10^{15}$ and $2 \times 10^{16}$ photon/cm$^{2}$. These yields were computed from our fixed energy experiments by applying the method described in section 2. For the binary mixed ices, these yields were corrected by dilution factors as explained before. In the following, we list important remarks concerning these results :
\\\\
- $^{12}$CO and $^{13}$CO are the most strongly desorbing species in each ice with a photodesorption yield at 564 eV from $\sim$ 0.1 to 0.8 molecule/photon. For $^{13}$CO:CH$_3$OH ice, we cannot distinguish HCO and $^{13}$CO molecules when looking at the photodesorption signal of mass 29. However, we can safely assume that the signal is dominated by $^{13}$CO as it is the dominant molecule in the ice. In H$_2$O:CH$_3$OH ice, we found radical HCO photodesorption in the mass channel 29 at all fixed energies we tested during the irradiations.
\\
- $^{12}$CO$_2$ and $^{13}$CO$_2$ are the second most strongly desorbed species from our ices.  
\\
- as explained in the next section, we assumed that no photodesorption of radical CH$_2$OH or CH$_3$O from $^{13}$CO:CH$_3$OH ice occurs. However, we detected a clear photodesorption signal of CH$_2$OH and/or CH$_3$O from water-ice mixtures at all fixed energies we tested, which does not appear to come from the cracking of CH$_3$OH in the QMS as no desorption signal has been found on the mass 32.
\\
- OH photodesorption from H$_2$O:CH$_3$OH ice is not detected during irradiations. Regarding $^{13}$CO:CH$_3$OH ice, we were not able to correct the mass signal 17 for the cracking of possible photodesorbed H$_2$O into OH in the QMS as we did not measure the mass signal 18 (the correction factor would be $\sim$21\% of the mass signal 18). If photodesorption occurs on the mass 17 for $^{13}$CO:CH$_3$OH ice, we can assume that the mass signal 17 is dominated by $^{13}$CH$_4$ photodesorption because for pure methanol ice and H$_2$O:CH$_3$OH ice, no OH photodesorption takes place and CH$_4$ photodesorption might occur.
\\
- for the desorption signal on mass 16, we were not able to distinguish between $^{12}$CH$_4$, atomic O, and $^{13}$CH$_3$ photodesorption.
\\
- for the mass 15, we give the raw data that are not corrected for any cracking pattern (especially from the mass 16, which could be attributed to CH$_4$ photodesorption) and could be overestimated.
\\\\
These results are consistent with the X-ray irradiation experiments (250-1250 eV) of H$_2$O:CO:NH$_3$ (2:1:1) ice, covered by a layer of CO:CH$_3$OH (3:1), conducted by \citealt{ciaravella_x-ray_2020}, in the sense that the dominant photodesorbing molecules detected in our experiments are also CO, CO$_2$, HCO (only clearly identified and measured for H$_2$O:CH$_3$OH mixed ice), and H$_2$CO. Nevertheless, quantification of the photodesorption yields was not provided by Ciaravella and colleagues, and additional important results arise from our measurements and are discussed in the remaining paper. 

\begin{figure} [h!]
\centering
\includegraphics[width=8.5cm]{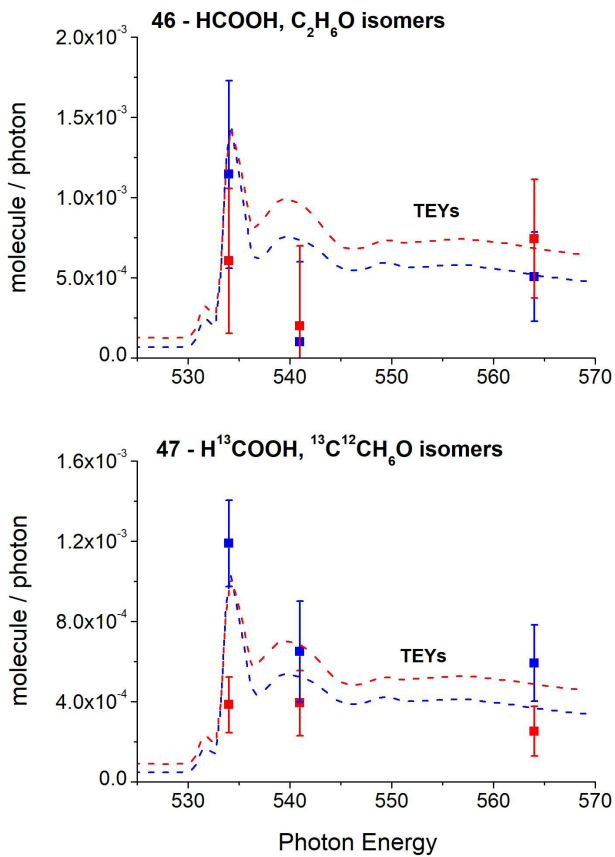}
\caption{Photodesorption spectra for masses 46 and 47 in molecule/photon from $^{13}$CO:CH$_3$OH ice at 15 K. Red and blue are associated with a mixing ratio of 1:1 and 6:1, respectively.}
\label{Graph_46_47_13CO}
\end{figure}

\subsubsection*{Organic molecules}

In Table \ref{Yield_COMS} we present the photodesorption yields of methanol (CH$_3$OH), formaldehyde (H$_2$CO), formic acid (HCOOH), and/or C$_2$H$_6$O isomers at 564 eV and for a fluence between $5 \times 10^{15}$ and $2 \times 10^{16}$ photon/cm$^{2}$. These yields were computed from our fixed energy experiments by applying the method described in Section 2. For mixture ices, these yields were corrected for dilution factors as explained before. In addition to the data reported in Table \ref{Yield_COMS}, we measured the signals for masses 60, 61, and 62 from $^{13}$CO:CH$_3$OH and H$_2$O:CH$_3$OH ices, but we did not detect any photodesorption (regarding the noise level on these channels, if photodesorption takes place, the corresponding yield is $<10^{-4}$ molecule/photon). These molecules could correspond to glycolaldehyde (HC(O)CH$_2$OH) or ethylene glycol (HOCH$_2$CH$_2$OH), which have been detected as a possible product of radical-radical recombination between HCO$\bullet$ and $\bullet$CH$_2$OH in VUV irradiated formaldehyde ice \citep{butscher_formation_2015, butscher_radical_2017}. 
\\\\
For $^{13}$CO:CH$_3$OH ice, we made the same assumption as for pure methanol ice to derive the X-ray photodesorption of CH$_3$OH, which is to consider the signal on the mass channel 31 as only originating from the cracking of desorbing CH$_3$OH into CH$_2$OH or CH$_3$O (which are the main fragments) in the ionization chamber of the QMS, neglecting the possible desorption of CH$_2$OH or CH$_3$O radical (which would contribute to the mass channel 31, see paper I for more details). This method seems less robust for $^{13}$CO:CH$_3$OH ice as the signal on the mass channel 31 could also correspond to H$_2^{13}$CO photodesorption and the signal on the mass 32 could also correspond to $^{13}$CH$_2$OH or $^{13}$CH$_3$O. At 564 eV, for a fresh ice (fluence <5.10$^{15}$ photon/cm$^2$) and for a mixing ratio of 1:1, we found that a maximum of $\sim$80\% of the mass signal 32 could correspond to CH$_3$OH photodesorption from $^{13}$CO:CH$_3$OH ice. This brings the mass signal 31 at 564 eV to below our detection threshold after correction for the cracking of CH$_3$OH into CH$_2$OH or CH$_3$O for this mixing ratio. For a mixing ratio of 6:1 and also for a fresh ice after it is corrected for the cracking of the entire mass 32 signal, some signal remained on the mass 31 at 564 eV and was attributed to H$_2^{13}$CO photodesorption, which is expected to be higher when more $^{13}$CO molecules are present in the binary mixed ice. Finally, X-ray photodesorption of the mass 33 from $^{13}$CO:CH$_3$OH ice, which unambiguously corresponds to methanol $^{13}$CH$_3$OH desorption, is also detected. These results are summarized in Table \ref{Yield_COMS}. 
\\\\
We also observed the photodesorption of masses higher than 32 for $^{13}$CO:CH$_3$OH mixtures. The photodesorption signal on the mass 47 can come from different phenomena: it can originate in the photodesorption of H$^{13}$COOH (formic acid) and/or $^{13}$C$^{12}$CH$_6$O isomers (ethanol and dimethyl ether), or it can come from the cracking of the the mass 48 ($^{13}$C$_2$H$_6$O isomers) in the QMS. These molecules have been detected by infrared spectroscopy when irradiating pure methanol ice at 14 K with X-rays of 550 eV \citep{chen_soft_2013}. As we did not measure the signal on the mass 48, we are not able to clearly identify the origin of the mass signal 47, and thus we did not correct the mass signal 46 for any possible cracking of the mass 47 in the QMS. In Figure \ref{Graph_46_47_13CO} we display the mass signals 46 and 47 from $^{13}$CO:CH$_3$OH mixture experiments without correction for the dilution factor and any cracking pattern. In the case of H$_2$O:CH$_3$OH ices, no X-ray photodesorption of HCOOH and/or C$_2$H$_6$O isomers were detected.

\section{Discussion}
Before we discuss the results derived for our binary mixed ice experiments, we recall the main conclusions regarding our results for pure methanol ice from paper I. The correlation between the photodesorption spectra and the TEY and the comparison between UV (from \citep{bertin_uv_2016}) and X-ray photodesorption yields led us to conclude that X-ray induced electron-stimulated desorption (XESD), dominated by low-energy secondary electrons resulting from the thermalization of the Auger electron after X-ray absorption by the ice, is the dominant process explaining the photodesorption yields from pure methanol ice. Moreover, we concluded that X-ray induced chemistry, also dominated by low-energy secondary electrons, plays an important role in X-ray irradiated methanol-containing ices. These conclusions are supported by the following discussion of binary mixed ices.

\subsection{X-ray photodesorption from $^{13}$CO:CH$_3$OH mixtures}
In $^{13}$CO:CH$_3$OH mixtures, TEY and the photodesorption yield are also strongly correlated (see Figure \ref{Graph_13CO}). Near 534 eV, photodesorption is triggered by the X-ray absorption of $^{13}$CO (or $^{12}$CO) molecules, whereas near 541 eV, it is triggered by the X-ray absorption of CH$_3$OH. Compared to pure methanol ice, mixing CH$_3$OH and $^{13}$CO molecules in condensed phase should not bring additional reaction channels in the X-ray induced chemistry network because CO, C, and O are already core species of the reaction network in pure methanol ice. However, the dynamics of the chemistry could be different as the initial conditions differ: $^{13}$CO molecules are already available in significant quantity when $^{13}$CO:CH$_3$OH mixtures are irradiated. \\

\begin{figure} [h!]
\resizebox{\hsize}{!}{\includegraphics{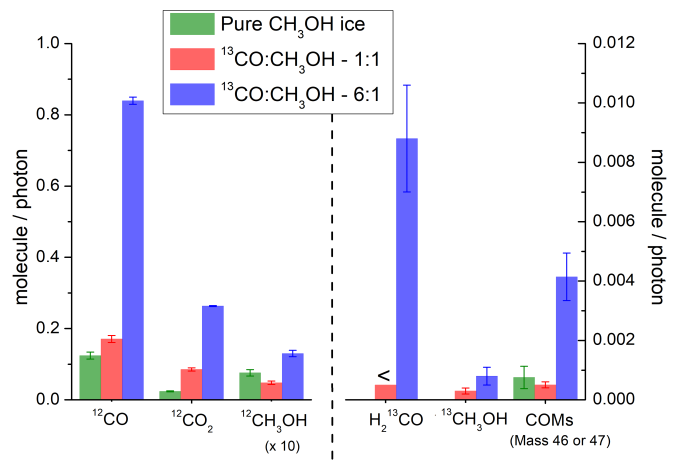}}
\caption{X-ray photodesorption yields at 564 eV of simple species and products from pure methanol ice (from paper I) and from $^{13}$CO:CH$_3$OH ice. These yields are taken from Tables \ref{Yields_Simple_Mol} and \ref{Yield_COMS}. In the right panel, COMs designate the X-ray photodesorption of the mass 46 (HCOOH and/or C$_2$H$_6$O) from pure methanol ice or the mass 47 (H$^{13}$COOH and/or $^{13}$C$^{12}$CH$_6$O) from $^{13}$CO:CH$_3$OH ice.}
\label{Pur_vs_mix_13CO}
\end{figure}

In Figure \ref{Pur_vs_mix_13CO} we quantitatively compare the X-ray photodesorption yields at 564 eV from Tables \ref{Yields_Simple_Mol} and \ref{Yield_COMS} (that are corrected for dilution factors as explained in section 3.2) between pure methanol ice (from paper I) and $^{13}$CO:CH$_3$OH mixtures, as a function of the mixing ratio. First, we should note that at 564 eV, the TEY value is approximately the same for pure methanol ice and for $^{13}$CO:CH$_3$OH ice, that is, $\sim 0.014$ electron/photon. Thus, the differences observed in the photodesorption yields between these ices should not be due to a difference in the efficiency of low-energy electron production per photon. In the following, we highlight important remarks, according to Figure \ref{Pur_vs_mix_13CO}, to discuss the possible mechanisms at play in X-ray irradiated $^{13}$CO:CH$_3$OH ice:
\\\\
- $^{12}$CO photodesorption efficiency is increased by almost one order of magnitude when methanol is sufficiently diluted in $^{13}$CO ice, from 0.12 molecule/photon or 0.17 molecule/photon for pure methanol ice and $^{13}$CO:CH$_3$OH (1:1) ice, respectively, to 0.84 molecule/photon for $^{13}$CO:CH$_3$OH (6:1) ice. This increase can only be due to interactions between $^{13}$CO and CH$_3$OH molecules during irradiations as induced chemistry involving only $^{13}$CO molecules cannot result in the formation of $^{12}$CO.
\\\\
- $^{12}$CO$_2$ photodesorption efficiency is also increased by one order of magnitude when methanol is sufficiently diluted in $^{13}$CO ice : $^{12}$CO$_2$ photodesorbed from pure methanol ice with a yield of 0.024 molecule/photon, whereas it photodesorbed from $^{13}$CO:CH$_3$OH (6:1) ice with a yield of 0.26 molecule/photon.
\\\\
- when we consider the effect of the mixing ratio in $^{13}$CO:CH$_3$OH mixtures alone (right panel of Figure \ref{Pur_vs_mix_13CO}), we observe an increase in the photodesorption efficiency of H-bearing $^{13}$CO molecules such as $^{13}$CH$_3$OH, H$^{13}$COOH and/or $^{13}$C$^{12}$CH$_6$O isomers (mass 47), when the dilution factor of methanol is increased. Although the estimate made for H$_2^{13}$CO is less robust, we also clearly see this trend for its photodesorption.
\\\\
These results show that mixing methanol with CO molecules in condensed phase could have a significant effect on the X-ray photodesorption, which also seems to be driven by X-ray induced-chemistry. When CH$_3$OH molecules are surrounded by enough $^{13}$CO molecules in condensed phase, the X-ray induced dissociation of CH$_3$OH into $^{12}$CO, most probably by H abstraction (as suggested in kinetic modeling of 5 keV electron irradiated methanol ice by \citealt{bennett_mechanistical_2007}), seems to be favored compared to the case of pure methanol ice and increases $^{12}$CO desorption. Consequently, $^{12}$CO$_2$ formation and subsequent desorption is mechanically increased by additional formation reaction channels between $^{12}$CO and $^{13}$CO molecules in $^{13}$CO:CH$_3$OH ice compared to the case of pure methanol ice. Of the possible additional reaction channels that could occur in our experiments, reaction (1) between electronically excited $^{13}$CO (or $^{12}$CO) and $^{12}$CO (or $^{13}$CO) has been suggested as a possible channel to form CO$_2$ in irradiation experiments of pure CO ice by 5 keV electrons \citep{jamieson_understanding_2006}, by Lyman-$\alpha$ photon \citep{Gerakines_1996}, and by 200 keV protons \citep{loeffler_comathsf_2_2005}. Finally, in our X-ray irradiation experiments, the remaining products resulting from the dissociation of CH$_3$OH into CO, which are most probably mainly H and/or H$_2$ (as suggested in \citealt{bennett_mechanistical_2007}), should react with the surrounding $^{13}$CO molecules in the case of $^{13}$CO:CH$_3$OH ice to form H-bearing $^{13}$CO molecules such as H$_2^{13}$CO, $^{13}$CH$_3$OH, H$^{13}$COOH and/or $^{13}$C$^{12}$CH$_6$O isomers. These effects seem to be indirectly probed by our photodesorption measurements when the dilution of CH$_3$OH in $^{13}$CO ice is increased,
\reaction{^{13 / 12}CO^{*}   +   ^{12 / 13}CO ->  ^{12}CO2  +  ^{13}C}
Our results also suggest that the X-ray photodesorption efficiency of formic acid, ethanol, and/or dimethyl ether is increased in CO-rich/CH$_3$OH-poor ice compared to pure methanol ice, as shown in Figure \ref{Pur_vs_mix_13CO}. Here we compare the photodesorption yield at 564 eV of the mass channel 46 from pure methanol ice with the photodesorption yield of the mass channel 47 from $^{13}$CO:CH$_3$OH ice, the latter being four times higher for a mixing ratio of 6:1. The direct availability of $^{13}$CO molecules to react with CH$_3$OH and/or its fragments could explain this result.
\\\\
Finally, we do not see a significant change in the estimated yields for CH$_3$OH photodesorption from $^{13}$CO:CH$_3$OH mixtures compared to pure methanol ice. However, as explained in section 3.2., these estimated yields are less robust than those derived for pure methanol ice, which makes the comparison unreliable. Our previous discussion of the effect of diluting CH$_3$OH in CO ice is not in favor of an efficient X-ray photodesorption of methanol when it is mixed in CO-rich ice, and additional X-rays experiments should be conducted to better constrain this yield for higher dilution factors of methanol in CO ice. However, $^{13}$CH$_3$OH (mass 33) X-ray photodesorption from our $^{13}$CO:CH$_3$OH ice (for both mixing ratios of 1:1 and 6:1), resulting from the X-ray induced-chemistry between $^{13}$CO and CH$_3$OH molecules and/or its dissociation products, is unambiguously detected and appears to be more efficient when the dilution factor of CH$_3$OH is increased, although it is less efficient by one order of magnitude than methanol X-ray photodesorption from pure methanol ice, which is assumed to be due to CH$_2$OH and/or CH$_3$O recombination, as discussed in paper I.
\\\\
The previous quantitative comparisons indicate that X-ray induced-chemistry could be intimately linked to the photodesorption process from CO:CH$_3$OH binary mixed ice. The initial conditions (i.e., the mixing ratio) appears to play a main role, especially for the X-ray photodesorption of methanol and some COMs such as formic acid, dimethyl ether, or ethanol. 

\subsection{X-ray photodesorption from H$_2$O:CH$_3$OH mixtures}
\begin{figure} [h!]
\resizebox{\hsize}{!}{\includegraphics{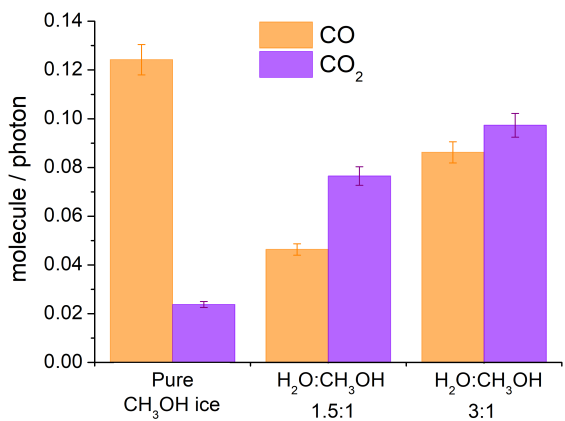}}
\caption{X-ray photodesorption yields at 564 eV of CO and CO$_2$ from pure methanol ice (from Part I) and from H$_2$O:CH$_3$OH ice. These yields are taken from Tables \ref{Yields_Simple_Mol} and \ref{Yield_COMS}.}
\label{Pur_vs_mix_H2O}
\end{figure}

For H$_2$O:CH$_3$OH mixtures, X-ray induced chemistry may play a main role in the photodesorption process. \citealt{laffon_photochemistry_2010} estimated based on NEXAFS spectroscopy that X-ray irradiation at 150 eV of H$_2$O:CH$_3$OH ice at 20 K leads to a methanol survival rate of $\sim$45\% and $\sim$25\% for 1:1 and 84:16$\sim$5:1 mixture ratios, respectively, after a dose of 1.1 MGy. \citealt{laffon_photochemistry_2010} assumed that this decrease in the survival rate of methanol when the H$_2$O concentration is increased was mainly explained by destruction of methanol reacting with the OH radical (this is condensed in reaction (3)), where OH radical comes from the dissociation of H$_2$O by photolysis and/or radiolysis via reaction (2). \citealt{laffon_photochemistry_2010} also observed that CO formation via reaction (3) and OH radical production via water dissociation (2) enhances CO$_2$ formation via reaction (4) in H$_2$O:CH$_3$OH ice compared to pure methanol ice, 
\begin{flalign}
& \ce{H_2O \xrightarrow{(h\nu, e^{-})} OH^{.} + H} \\
& \ce{CH3OH + 4 OH^{.} -> CO + 4 H2O} \\
& \ce{CO + OH^{.} -> CO2 + H}
\end{flalign}
\\
Water dissociation by channel (2) is the main dissociation channel in X-ray irradiated pure water ice \citep{Laffon_2006, laffon_photochemistry_2010}. In our experiments on H$_2$O:CH$_3$OH mixtures, radical OH production is then expected to be significant, and the survival rate of CH$_3$OH should be very low. Our results also agree well with those of \citealt{laffon_photochemistry_2010}: in Figure \ref{Pur_vs_mix_H2O},  we see that CO yield is higher than CO$_2$ yield for pure methanol ice by almost one order of magnitude at 564 eV, whereas in H$_2$O:CH$_3$OH ice (for both mixing ratios of 1.5:1 and 3:1), the CO yield is lower than the CO$_2$ yield. Consumption of OH radical by reaction (3) and (4) may also explain why we do not detect an OH radical contribution to the TEY in H$_2$O:CH$_3$OH ice compared to pure H$_2$O ice (see Figure \ref{TEY}) for which fewer consumption channels of the OH radical are available. \\

OH reactivity with CH$_3$OH at low temperature (T < 100 K) in gas phase has been proven to be very efficient due to possible quantum tunneling effect \citep{gomez_martin_low_2014, ocana_gas-phase_2019}. The main dissociation product was proposed to be CH$_3$O \citep{shannon_accelerated_2013}. A similar behavior is expected in condensed phase and could even be enhanced as activation barriers of reactions are strongly reduced and reaction probabilities are increased compared to gas-phase reactions.  Thus, radical OH production may play a main role in X-ray irradiated H$_2$O:CH$_3$OH ice by opening a new reaction channel to destroy CH$_3$OH molecules compared to the case of pure methanol ice. This may explain why we do not observe any photodesorption of CH$_3$OH (mass 32) from H$_2$O:CH$_3$OH ice, whereas a clear photodesorption signal of CH$_2$OH and/or CH$_3$O (mass 31) is detected. Finally, the nondetection of a photodesorption signal on the mass 46 seems to indicate that mixing water with methanol molecules may close reaction channels that lead to the formation of COMs in the X-ray induced chemistry network.

\subsection{Astrophysical implications}
In paper I, we have extrapolated our experimental photodesorption yields from pure methanol ice to the protoplanetary disk environment and showed that X-ray photodesorption is an efficient process (at least as efficient as UV photodesorption) to desorb molecules from pure methanol ices. However, our results on binary mixed ices (section 3) show that photodesorption is strongly dependent on the ice composition so that the case of pure methanol ice is not necessarily astrophysically relevant. In the cold regions of protoplanetary disks, the ice composition is different depending on the distance from the young star. At temperatures between 30 and 77 K, H$_2$O-rich ices are dominant, with some possible traces of CH$_3$OH \citep{boogert_observations_2015}. Beyond the CO snow line, for T < 20 K, interstellar ices are expected to be composed of an inner H$_2$O-rich layer on top of which a CO-rich layer containing CH$_3$OH is formed. Our experimental results indicate that the possible X-ray photodesorption of molecules from interstellar ices should produce different outcomes depending on the region considered: in the H$_2$O-rich ice cold regions, X-ray photodesorption should enrich the gas phase with simple molecules mostly such as CO or CO$_2$ , whereas in the top layer, CO-rich ice-cold regions, in addition to simple molecules, COMs such as methanol, formic acid, ethanol, and/or dimethyl ether should also be ejected in the gas phase. In the H$_2$O-rich ice regions, X-ray photodesorption of radicals such as HCO$\bullet$ or $\bullet$CH$_2$OH/$\bullet$CH$_3$O could open new reaction pathways in the gas phase to produce more complex molecules. The possible X-ray photodesorption of ethanol from CO:CH$_3$OH ices is also very interesting as gas-phase reactions involving ethanol can lead to more complex molecules such as glycolaldehyde, acetic acid, and formic acid \citep{Skouteris_2018}. However, we should note that the methanol dilution factors used in our experiments could differ from what is observed in protoplanetary disks and that its effect on the X-ray photodesorption from methanol-containing ices remains to be better constrained. \\
\begin{figure} [h!]
\centering
\includegraphics[width=8cm]{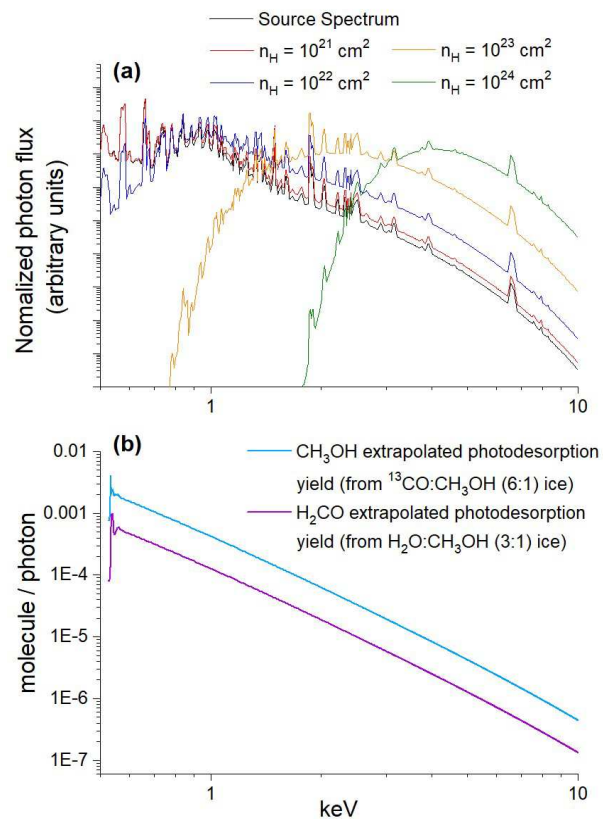}
\caption{(a) Normalized (with respect to the area) X-ray spectra of TW Hya from \citealt{nomura_molecular_2007} (source spectrum and its attenuation for different H column densities). (b) Extrapolated photodesorption yield spectra of CH$_3$OH from $^{13}$CO:CH$_3$OH (6:1) ice and of H$_2$CO from H$_2$O:CH$_3$OH (3:1) ice.}
\label{Astro_II}
\end{figure}

\begin{table}[!h]
\caption{Coefficient to apply to the yields in Table \ref{Astro_Yields_Comp} (multiplication) in order to obtain the astrophysical yields corresponding to other attenuation factors than $n_H = 10^{23}$ cm$^2$.}

\begin{center}
\begin{threeparttable}

\begin{tabular}{p{3cm}p{3cm}}

\hline \hline \\
\centering Attenuation factor & \multicolumn{1}{c}{Coefficient} \\[0,2cm]\hline\\

\centering $n_H = 0$ cm$^2$ & \multicolumn{1}{c}{16.8}  \\[0,1cm] 

\centering $n_H = 10^{21}$ cm$^2$ & \multicolumn{1}{c}{15.3}  \\[0,1cm] 

\centering $n_H = 10^{22}$ cm$^2$ & \multicolumn{1}{c}{7.66}  \\[0,1cm]

\centering $n_H = 10^{23}$ cm$^2$ & \multicolumn{1}{c}{1}  \\[0,1cm]  

\centering $n_H = 10^{24}$ cm$^2$ & \multicolumn{1}{c}{0.13} \\[0,1cm]  

\hline
          
\end{tabular}   
\end{threeparttable}
\end{center}   
\label{att_factor}

\end{table}

\begin{table*}[h!]
\caption{Average astrophysical photodesorption yield in molecule/photon extrapolated from our experimental results, using the method described in section 4.3 (for more details, see paper I), for different molecules and different ice mixtures at 15 K and without correction for any dilution factor. The X-ray emission spectrum used is that of TW Hya from \citealt{nomura_molecular_2007}, to which we applied an attenuation factor corresponding to $n_H = 10^{23}$ cm$^2$.}

\begin{center}
\begin{threeparttable}

\begin{tabular}{p{3cm}p{3cm}p{3cm}p{3cm}p{3cm}}

\hline \hline \\
\multirow{2}{3cm}{\centering Desorbed \\ molecule} & \multicolumn{2}{c}{\multirow{1}{*}{CO:CH$_3$OH ice}} &
\multicolumn{2}{c}{\multirow{1}{*}{H$_2$O:CH$_3$OH ice}}
\\\cmidrule{2-3}\cmidrule{4-5}

& \centering 6:1 & \centering 1:1 & \centering 3:1 & \multicolumn{1}{c} {1.5:1} 
\\[0,1cm]\hline\\

\centering H$_2$O &  & &  \centering 5.1 $\pm$ 2.6 $\times 10^{-5}$ &  \multicolumn{1}{c} {6.2 $\pm$ 3.1 $\times 10^{-5}$} \\[0,1cm] 

\centering CO  &  \centering 1.5 $\pm$ 0.8 $\times 10^{-2}$ &  \centering 6.4 $\pm$ 3.2 $\times 10^{-3}$ & \centering 6.5 $\pm$ 3.3 $\times 10^{-4}$ &  \multicolumn{1}{c} {5.6 $\pm$ 2.7 $\times 10^{-4}$} \\[0,1cm]
 
\centering CO$_2$  &  \centering 3.3 $\pm$ 1.7 $\times 10^{-3}$ &  \centering 2.2 $\pm$ 1.1 $\times 10^{-3}$ &  \centering 7.3 $\pm$ 3.7 $\times 10^{-4}$ &  \multicolumn{1}{c} {9.2 $\pm$ 4.6 $\times 10^{-4}$} \\[0,1cm]
  
\centering HCO & & &  \centering 6.3 $\pm$ 3.2 $\times 10^{-4}$ &  \multicolumn{1}{c} {1.9 $\pm$ 0.9 $\times 10^{-4}$} \\[0,1cm]

\centering H$_2$CO  &  \centering 3.9 $\pm$ 2.0 $\times 10^{-5}$ &  &  \centering 1.7 $\pm$ 0.9 $\times 10^{-5}$ &  \multicolumn{1}{c} {1.4 $\pm$ 0.7 $\times 10^{-5}$} \\[0,1cm]

\centering CH$_2$OH, CH$_3$O  &  & &  \centering 3.9 $\pm$ 2.0 $\times 10^{-5}$ &  \multicolumn{1}{c} {1.5 $\pm$ 0.7 $\times 10^{-5}$}\\[0,1cm]

\centering CH$_3$OH  &  \centering $[3.4 \pm 1.7 \times 10^{-6}$; &  \multicolumn{1}{c} {$[4.0 \pm 2.0 \times 10^{-6}$;} &  & \\[0,1cm]

 &  \centering $5.7 \pm 2.4 \times 10^{-5}]$\tnote{*} &  \multicolumn{1}{c} {$7.5 \pm 3.7 \times 10^{-5}]$\tnote{*}} & & \\[0,1cm]

\centering HCOOH, C$_2$H$_6$O  &  \centering 1.8 $\pm$ 0.9 $\times 10^{-5}$ &  \multicolumn{1}{c} {7.8 $\pm$ 3.9 $\times 10^{-6}$} & & \\[0,1cm]

\hline
          
\end{tabular}
\begin{tablenotes}
            \item[*] for CH$_3$OH X-ray photodesorption from CO:CH$_3$OH ice, we provide a range of value extrapolated from our experimental yields of  $^{13}$CH$_3$OH (minimum value) or $^{12}$CH$_3$OH (maximum value) desorption from $^{13}$CO:CH$_3$OH ice, the latter estimated yield is less robust, as explained in section 3.2.
\end{tablenotes}

\end{threeparttable}
\end{center}   
\label{Astro_Yields_Comp}

\end{table*}
To take the differences induced by the ice composition on X-ray photodesorption into account, we extrapolated our experimental photodesorption yields from our binary mixed ices to the protoplanetary disk environment using the same method as in paper I: we computed the local X-ray field (normalized) corresponding to the attenuated emission spectrum of a Classical T Tauri star (TW Hya from \citealt{nomura_molecular_2007}), depending on the H column density considered \citep{bethell_photoelectric_2011}, and we multiplied it with our extrapolated photodesorption spectra, which are assumed to follow the TEY of the considered binary mixed ice between 0.525 keV and 0.570 keV (starting from the estimated yield at 564 eV from Table \ref{Yields_Simple_Mol} or \ref{Yield_COMS}, without correction for any dilution factor) and are extrapolated above 0.570 keV according to the X-ray absorption cross section of gas-phase methanol \citep{Berkowitz:1087021}. In Figure \ref{Astro_II}.(a) we display the local X-ray field we computed, and in Figure \ref{Astro_II}.(b) we display an example of the extrapolated photodesoption spectra: one of CH$_3$OH from $^{13}$CO:CH$_3$OH (6:1) ice, and one of H$_2$CO from H$_2$O:CH$_3$OH (3:1) ice. The final computations are provided in Table \ref{Astro_Yields_Comp} and represent astrophysical X-ray photodesorption yields that could be used for astrochemical modeling. For the specific case of methanol X-ray photodesorption from CO:CH$_3$OH, we provide a range of values extrapolated from our experimental yields by considering whether the yield of $^{13}$CH$_3$OH (minimum value) or $^{12}$CH$_3$OH (maximum value) desorption from $^{13}$CO:CH$_3$OH ice as the latter estimated yield is less robust, as explained in section 3.2. To compute the astrophysical yields corresponding to other attenuation factors, the coefficients displayed in Table \ref{att_factor} should be applied (multiplication) to the yields of the Table \ref{Astro_Yields_Comp}. As already mentioned in sections 4.1 and 4.2, X-ray photodesorption depends on the ice composition and on the nature of the desorbing species but also the mixing stoichiometry. When possible, these parameters could be included in astrochemical models.

\section{Conclusion}
$^{13}$CO:CH$_3$OH ice and H$_2$O:CH$_3$OH ice were irradiated by monochromatic X-rays in the range of 525-570 eV. Intact methanol, other COMs, and simpler molecules were found to photodesorb due to X-ray absorption of core O(1s) electrons, quantified via TEY measurement, which leads to a cascade of low-energy secondary electrons within the ice. X-ray photodesorption yields were derived and found to be intimately linked to X-ray induced chemistry, which indicates that X-ray induced electron-stimulated desorption (XESD) may be the dominant mechanism explaining X-ray photodesorption from these ices. However, electron-stimulated desorption experiments are mandatory to conclude on the dominant mechanism. The main conclusions of this paper are listed as follows:

\begin{enumerate}

\item In $^{13}$CO:$^{12}$CH$_3$OH ice, $^{12}$CH$_3$OH X-ray photodesorption is estimated to be $\sim 10^{-2}$ molecule desorbed by incident photons, although this estimation is less robust. $^{13}$CH$_3$OH desorption is unambiguously detected but is less efficient by one order of magnitude.

\item X-ray photodesorption of formic acid, ethanol, and/or dimethyl ether is detected with a yield of $\sim 10^{-3}$ molecule desorbed per incident photon for $^{13}$CO:$^{12}$CH$_3$OH ice. In the cold regions of protoplanetary disks, beyond the CO snowline, X-rays should then participate in the enrichment of the gas phase with these COMs, in addition to methanol.

\item When methanol is mixed in water mantles, the production of radical OH due to the dissociation of water molecules and its reaction with CH$_3$OH is assumed to play a dominant role in the X-ray induced chemistry. Consequently, X-ray photodesorption of methanol and previous COMs from H$_2$O:CH$_3$OH ice is not detected, but radicals such as HCO and CH$_2$OH/CH$_3$O are photodesorbing.  

\end{enumerate}
Astrophysical X-ray photodesorption yields of several species and for different ice composition, extrapolated from our experimental yields, are also provided in Table \ref{Astro_Yields_Comp} for astrochemical modeling.

\begin{acknowledgements}
This work was done with financial support from the Region Ile-de-France
DIM-ACAV+ program and by the European Organization for Nuclear Research
(CERN) under the collaboration Agreement No. KE3324/TE. We would like to
acknowledge SOLEIL for provision of synchrotron radiation facilities
under Project Nos. 20181140, and we thank N. Jaouen, H. Popescu and R.
Gaudemer for their help on the SEXTANTS beam line. This work was
supported by the Programme National “Physique et Chimie du Milieu
Interstellaire” (PCMI) of CNRS/INSU with INC/INP co-funded by CEA and
CNES. Financial support from the LabEx MiChem, part of the French state
funds managed by the ANR within the investissements d’avenir program
under Reference No. ANR-11-10EX-0004-02, is gratefully acknowledged.
\end{acknowledgements}

\bibliographystyle{aa}
\bibliography{Bibli}

\end{document}